\documentclass[12pt]{iopart}
\usepackage{amssymb,epsf}

\newcommand{\MeV}{\mbox{ MeV}}
\newcommand{\GeV}{\mbox{ GeV}}
\newcommand{\Msun}{\mbox{ M}_\odot}

\begin{document}
\jl{4}

\title{Possible Cosmological Implications of the Quark-Hadron Phase Transition}

\author{N Borghini\dag, W N Cottingham\ddag\ and R Vinh~Mau\dag}

\address{\dag\ Laboratoire de Physique Th\'eorique des Particules
El\'ementaires, Universit\'e Pierre et Marie Curie, 4 Place Jussieu 75252
Paris Cedex 05, France}

\address{\ddag\ Physics Department, University of Bristol, Bristol BS8 1TH, UK}

\begin{abstract}

We study the quark-hadron phase transition within an effective model of QCD,
and find that in a reasonable range of the main parameters of the model,
bodies with quark content between $10^{-2}$ and 10 solar masses can have
been formed in the early universe. 
In addition, we show that a significant amount of entropy is released during
the transition. 
This may imply the existence of a higher baryon number density than what is 
usually expected at temperatures above the QCD scale. 
The cosmological QCD transition may then provide a natural way for decreasing 
the high baryon asymmetry created by an Affleck-Dine like mechanism down to 
the value required by primordial nucleosynthesis. 

\end{abstract}


\submitted

\maketitle

\section{Introduction}

It has been shown that the quark-hadron phase transition which occurred in the 
early Universe could lead to the formation of relic quark-gluon plasma 
objects, which survive today \cite{witten84,applegate85}. 
Generally, it is admitted \cite{witten84,fuller,iso86,ng91} that the 
transition occurred effectively at the critical temperature, which is of the
order $100\MeV$, the QCD energy scale. 
In that case, the quark content of the bodies which have been formed during 
the transition cannot be larger than $10^{-8}\Msun$. 
Another possibility arises if the transition was delayed for some time. 
It then becomes possible that the quark plasma objects formed at the end of 
the transition appeared at a temperature much lower than $T_c$, and more
massive bodies may have been produced \cite{wnc94,wnc98,goyal98}; these 
latter could then account for some fraction of the dark matter in the Universe.

In this work, we perform a detailed analysis of the phase transition within an 
effective model of QCD, and we show that for the same critical temperature as 
what is usually thought, we can obtain a high degree of supercooling, which 
allows the possible formation of large ``quark stars'' with masses ranging 
from $10^{-2}$ to $10\Msun$. 
We find that this is so because the Universe has grown exponentially during 
the quark-hadron transition, at a temperature $T \lesssim 100\MeV$, much lower 
than the temperature of classical inflationary models \cite{lyth99}. 
Moreover, we show that the exponential expansion is not balanced by so steep a 
drop of temperature, and thus increases significantly the total entropy of the 
Universe. 
This entropy production dilutes the density of any conserved or 
quasi-conserved quantity present before the transition, such as the baryon 
number. 
Therefore, our model requires a high value of these quantities at $T \gtrsim 
T_c$, as for example the large baryon asymmetry produced in some 
supersymmetric models \cite{affleck85}. 
This leads us to incorporate the quark chemical potential in our model, and to 
study the cosmological quark-hadron transition in the context of both high 
temperature and quark number density.

\section{The Model}

Except in the quenched approximation, the theory of strong interactions QCD is 
too difficult, computationally, to give a reliable description of the phase 
transition between the high temperature and density phase of the quark gluon 
plasma and the low temperature and density phase of quarks bound in hadrons. 
In fact, lattice calculations  \cite{ukawa98} do not seem able yet to really tackle 
the problem of finite quark densities. 
Because the light quarks have such a small mass compared with the transition 
temperature or chemical potential, we believe them to play an essential role 
in determining the quantitative features of the transition which is not taken 
into account in the quenched approximation, as shows for example the 
discrepancy between the values of the critical temperature computed with or 
without dynamical quarks. 
In considering the transition, we therefore work with a simple model in which 
non-perturbative QCD is replaced by an effective Lagrangian which incorporates 
the chiral symmetry of the theory. 
In this model the quark fields are interacting with a chiral field formed with 
the $\bi{\pi}$ meson field and the scalar field $\sigma$. 
The Lagrangian density is 
\begin{equation}
 \fl {\cal L} = \sum_{k=1}^{n_{\rm f}} \left[ i\bar{\psi}_k \gamma^\mu 
\partial_\mu \psi_k - g \bar{\psi}_k (\sigma + i \bi{\tau} \cdot \bi{\pi} 
\gamma_5) \psi_k \right] + \frac{1}{2} \partial_\mu \sigma \partial^\mu \sigma
+\frac{1}{2}\partial_\mu \bi{\pi} \partial^\mu \bi{\pi} - V(\sigma^2 + \pi^2)
\label{lagrangian1}
\end{equation}
which can be rewritten as \cite{kalafatis92}
\begin{equation}
\fl {\cal L} = \sum_{k=1}^{n_{\rm f}} \left[ i\bar{\psi}_k \gamma^\mu 
\partial_\mu \psi_k - g \xi (\bar{\psi}_k^{\rm L} U \psi_k^{\rm R} + 
\bar{\psi}_k^{\rm R} U^+ \psi_k^{\rm L}) \right] + \frac{1}{2} \partial_\mu 
\xi \partial^\mu \xi +\frac{1}{4} \xi^2\ \Tr (\partial_\mu U \partial^\mu U^+) 
- V(\xi)
\label{lagrangian2}
\end{equation}
where $\psi_k^{\rm L,R}$ are the left- and right-handed components of the quark
field $\psi_k$, $U$ is an element of SU(2) defined by $\xi U = \sigma + i 
\bi{\tau} \cdot \bi{\pi}$ and $\xi = (\sigma^2 + \pi^2)^\frac{1}{2}$.

The generalized self-interaction potential is the usual quartic function in
$\xi$. We choose to write it in a seemingly complicated form, but such that
the parameters $f_\pi$, $\lambda$, $B$ are readily related to physical
quantities.
\begin{equation}
\fl V(\xi)=\frac{1}{2} f_\pi^2 \left(\lambda^2  - \frac{12B}{f_\pi^4}\right)
\xi^2 \left(1  - \frac{\xi}{f_\pi}\right)^2 + B \left[1 +
3\left(\frac{\xi}{f_\pi}\right)^4 - 4\left(\frac{\xi}{f_\pi}\right)^3 \right]
\label{eff_potbis}
\end{equation}
$V(\xi)$ has its absolute minimum at $\xi = f_\pi = 93\MeV$, a value chosen to 
fit the observed pion decay rate. For this value, which corresponds to the 
physical vacuum, chiral symmetry is spontaneously broken; however, if 
$\bi{\pi} = 0$ isospin symmetry is preserved, just as in nature.
If $B < \frac{\lambda^2 f_\pi^4}{12}$, the potential has a second local 
minimum at $\xi = 0$, corresponding to a chirally symmetric, metastable ``false 
vacuum'' with the energy density $B$. 
This is analogous to the perturbative vacuum of the MIT bag model \cite{MITbag}, 
and $B$ can be interpreted as the bag constant in that model.
Various versions of the previous model have often been used to study the low
energy hadron spectroscopy \cite{wilets87} or heavy ion collisions 
\cite{csernai95}, and phenomenological fits to light hadron properties give 
$B^{1/4}$ between $100$ and $200\MeV$ \cite{lee92}.
Small oscillations of $\xi$ about the minimum at $\xi = f_\pi$ correspond to a 
scalar particle of mass $m_\xi = \lambda f_\pi$ and three massless 
pseudoscalar pions. 
We can notice that the $\xi$ field is a chiral singlet, and the associated 
scalar particle may be interpreted as representing the condensate arising from 
non-linear interactions of the gluons (glueballs). 
As the lightest glueballs are believed to have a mass in the range $1.5$ to 
$1.7\GeV$ \cite{glueball,PDG98}, $\lambda$ is between $16$ and $19$.
The coupling constant $g$ gives mass to the quarks, and thus helps to break the 
chiral symmetry; following the fits to hadronic physics, we take it to be larger
than 10, so that quarks have an effective mass larger than $1\GeV$ in the 
physical vacuum: it is then energetically unfavourable for them to exist in
the phase with $\xi = f_\pi$. 
However, the actual value of $g$ is not important for the following 
calculations.

In order to study the transition which took place in the early Universe, at a 
temperature of the order $100\MeV$, we must implement finite temperature in 
our model. 
According to the common lore, it seems irrelevant to take into account a 
possible chemical potential, because the quark density in the cosmological context 
is too small to be significant (see e.g. \cite{weinberg72}, pp. 530-531). 
To this effect, we must add to the potential given by equation 
(\ref{eff_potbis}) contributions like \cite{dolan74,kapusta89}
\begin{eqnarray}
f_f(T) & = & -2T \int\!\frac{d^3k}{(2\pi)^3} \ln\left(1 + e^{-E(k)/T}\right), 
\label{free_en_ferm1} \\
E(k) & = & \sqrt{k^2 + g^2\xi^2} \nonumber
\end{eqnarray}
for each fermionic degree of freedom (spin, flavour, colour), and
\begin{eqnarray}
f_b(T) & = & T \int\!\frac{d^3k}{(2\pi)^3} \ln\left(1 -
e^{-E^\prime(k)/T}\right), 
\label{free_en_bos1} \\
E^\prime(k) & = & \sqrt{k^2 + m_\xi^2} \nonumber
\end{eqnarray}
for bosonic degrees of freedom. These terms come from the one loop
approximation and represent the effect of the thermal excitations of the
quark-antiquark pairs and of the $\xi$ field.
These latter have a mass $\left( \frac{\partial^2 V}{\partial\xi^2}
\right)^{1/2}$ larger than $m_\xi$ in both phases with $\xi = 0$ and $\xi = 
f_\pi$, and are strongly suppressed by the Boltzmann factor $e^{-E^\prime/T}$ 
at $T \lesssim 100\MeV$. 
The same arguments afford us to discard the excitations of $\rm q \bar{\rm q}$ 
pairs in the physical vacuum $\xi = f_\pi$, but not in the false vacuum $\xi = 
0$, where the contributions given by equation (\ref{free_en_ferm1}) lower the 
value of the effective potential. 
For $\xi$ close to 0, which implies $\frac{g\xi}{T} \ll 1$, the integral can 
be expanded as \cite{dolan74,wnc91}
\begin{equation}
f_f(T) = -\frac{7\pi^2}{360} T^4 + \frac{1}{24}g^2 \xi^2 T^2
\label{free_en_ferm1b}
\end{equation}
Actually, we have checked numerically that equation (\ref{free_en_ferm1b})
represents a very accurate approximation.
We include in our model the three light quarks. Each contribute for 3 degrees 
of freedom of colour, and 2 of spin, giving 18 degrees of freedom in all. 
The u and d quarks are considered massless, whereas for the strange quark we 
will keep the quadratic term of equation (\ref{free_en_ferm1b}) to take into 
account its finite mass.

With these modifications, the self-interaction potential at finite temperature 
and $\xi = 0$ can be rewritten
\begin{equation}
V_T(\xi) = B - \alpha_T T^4 + \gamma_T T^2
\label{eff_pot_T}
\end{equation}
where $\alpha_T =  \frac{7\pi^2}{20}$ and $\gamma_T = \frac{1}{4} m_s^2$,
$m_s = (60-170)\MeV$.\cite{PDG98}

\begin{figure}
\begin{center}
\epsfbox{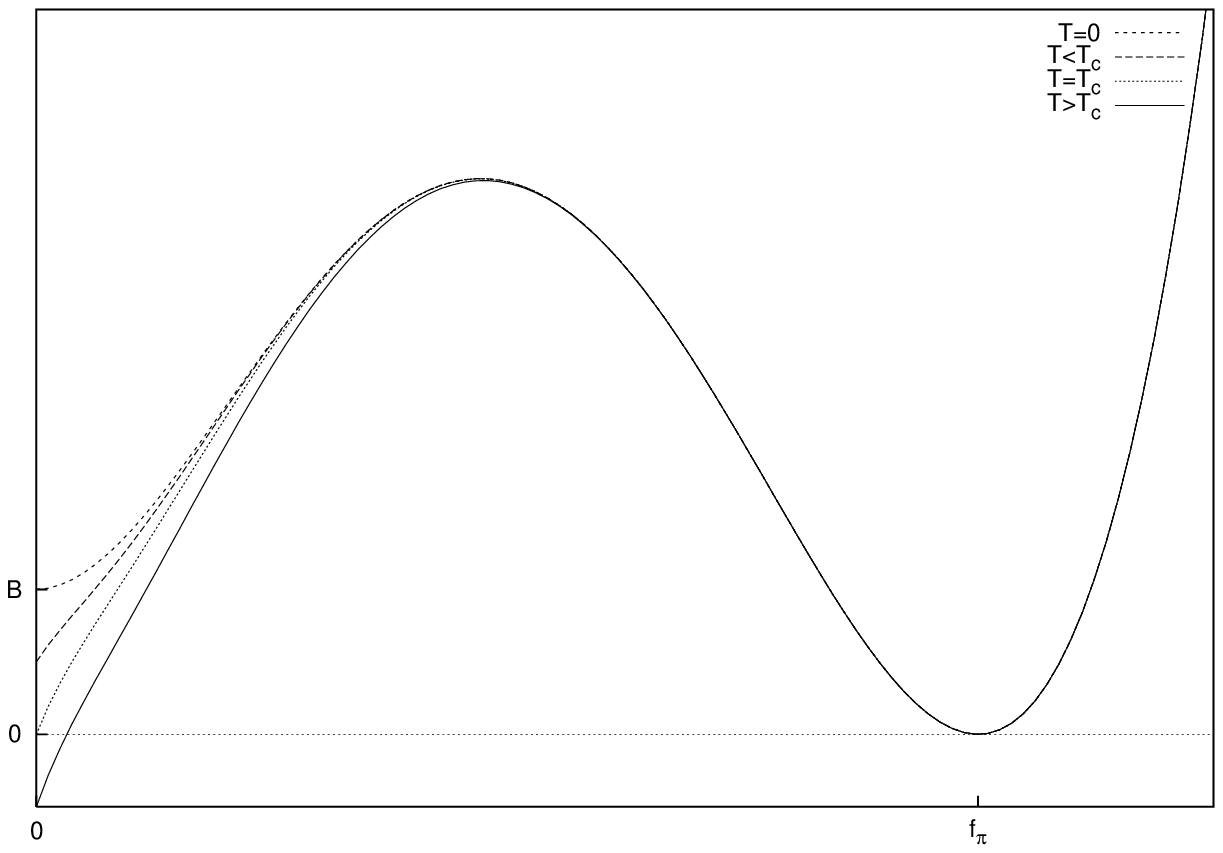}
\end{center}
\caption{The effective potential $V_T(x)$. 
Solid curve corresponds to the potential at zero temperature; 
dashed curves to the potential at $T < T_c$, $T = T_c$ and $T > T_c$.\label{plot-pot}}
\end{figure}

A sketch of $V_T(\xi)$ for different values of the temperature is given in
\fref{plot-pot}. 
It has two minima, at $\xi \simeq f_\pi$ with the same value as at zero 
temperature, and at $\xi = 0$, where the value of the potential is lowered. 
The overpressure, difference in free energy density between these minima, is
\begin{equation}
\Delta P = B + \gamma_T T^2 - \alpha_T T^4
\label{diff_P1}
\end{equation}
This difference vanishes when
\begin{equation}
T = T_c = \sqrt{\frac{\gamma_T + \sqrt{\gamma_T^2 + 4\alpha_T B}}{2\alpha_T}}
\label{Tc1}
\end{equation}
For this value, both minima of $V_T(\xi)$ are degenerate, which gives a first 
order phase transition. 
Even though the order of the QCD transition is still a highly debated issue, this 
is  in agreement with previous theoretical predictions using the linear $\sigma$ 
model \cite{pisarski84}, which, incidentally, is a particular case of our model 
when $\lambda^2 f_\pi^4 = 8B$, or with recent indications from lattice calculations 
with 3 degenerate Wilson quarks or with 2 massless quarks and a light s quark 
\cite{iwasaki96}. 
$T_c$ is then the critical temperature of the transition.
If $T>T_c$, the minimum at $\xi=0$ becomes the absolute minimum of $V_T(\xi)$.

Besides the quarks, we first include in our model a thermal bath of relativistic
particles which influence the transition only through gravitational effects: at 
this stage, they represent the ``physical vacuum''. 
The contribution to the free energy density of both phases of these spectator 
particles (photons, electrons, muons and neutrinos) is given by expressions 
(\ref{free_en_ferm1}) and (\ref{free_en_bos1}) for fermions with their 
antiparticles and bosons respectively. 
These terms can be expanded as $f_f = -\frac{7\pi^2}{360} T^4 + \frac{1}{24} 
m^2 T^2$, with a mass term $m_\mu$ only for the muon, and $f_b = 
-\frac{\pi^2}{90} T^4$ for the massless photon, so we get an extra contribution
\begin{equation}
f_v = -\frac{14.25\pi^2}{90} T^4 + \frac{1}{12} m_\mu^2 T^2 = -\alpha_v T^4 +
\gamma_vT^2
\label{p_v}
\end{equation}
contributing to both phases. 
However, if we consider the overpressure $\Delta P$ between the phases with
$\xi = 0$ and $\xi = f_\pi$, this extra free energy density cancels, and
$\Delta P$ is still given by equation (\ref{diff_P1}). 

In a more exhaustive calculation, we will also include in the hadronic phase $\xi = 
f_\pi$ the lightest hadrons, the pions, and show that their influence on the results, 
which we discuss later, is not qualitatively important.

\section{Bubble nucleation. Dynamics of the Universe}

\begin{figure}
\begin{center}
\epsfbox{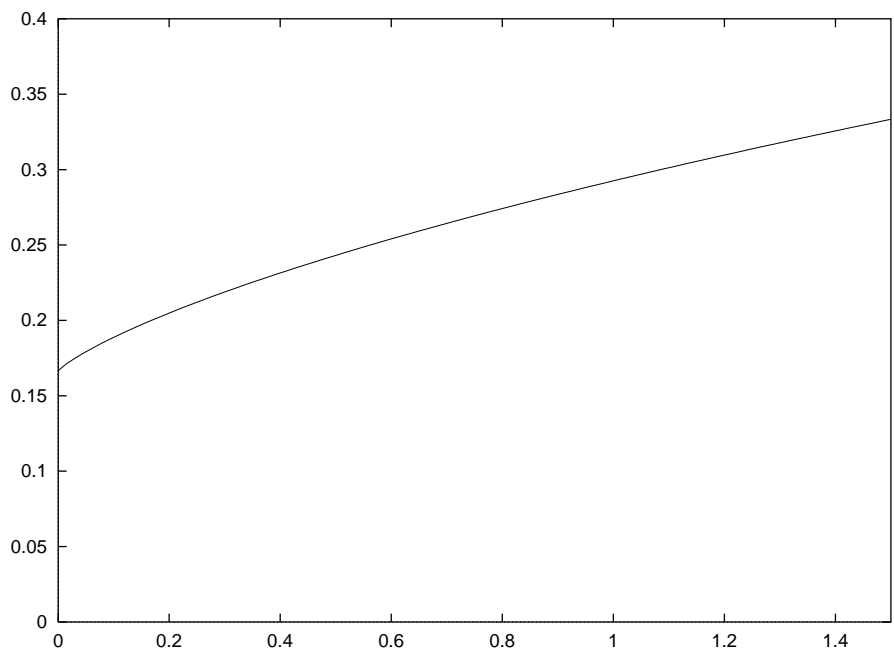}
\end{center}
\caption{Graph of the integral $I(y)$ of equation (\ref{action2}).\label{plot-I}}
\end{figure}

After the Big Bang, the temperature of the early Universe is higher than the
critical temperature of the quark-hadron transition (equation \ref{Tc1}): the
phase with $\xi = 0$ is the more stable. 
The massless quarks are deconfined -- this is the so-called quark-gluon plasma 
--, and chiral symmetry is preserved. 
According to the standard model of cosmology, the Universe expands, while its 
temperature decreases. 
When it drops below $T_c$, bubbles of the true vacuum $\xi = f_\pi$, which is 
now the stable phase, begin to appear within the quark plasma. 
If we consider only homogeneous nucleation, the nucleation rate per unit 
volume is given by \cite{linde83}
\begin{equation}
\Gamma(T) = C T^4 \left(\frac{S_3}{2\pi T} \right)^{3/2} e^{-S_3/T} 
\label{rate}
\end{equation}
$C$ is a multiplicative coefficient of the order unity \cite{csernai92}. 
For a spherical bubble, $S_3$ is the stationary value of the functional $F = 
4\pi \int r^2 \left[\frac{1}{2} (\frac{d\xi}{dr})^2 + V_T(\xi)\right]dr$. 
In the thin-wall approximation \cite{linde83}, this expression can be replaced 
by $F = -\frac{4}{3}\pi r_0^3 \Delta P + 4\pi r_0^2 s$, where $r_0$ is the 
bubble radius, the overpressure $\Delta P$ is the difference in $V_T(\xi)$ 
between the two minima, given by expression (\ref{diff_P1}), and $s$ is the 
surface tension.
This latter is 
\begin{equation}
s = \int_0^{f_\pi}\! \sqrt{2V_{T_c}(\xi)} d\xi = \lambda f_\pi^3 I(y)
\label{surf_tens}
\end{equation}
with \cite{wnc98}
\begin{equation}
I(y)=\int_0^1\ (1-u) \sqrt{u^2+\frac{y}{6}(1+2u-3u^2)}\ du\ ,\quad
y = \frac{12B}{\lambda^2 f_\pi^4}.
\label{action2}
\end{equation}
The graph of $I(y)$  in \fref{plot-I} shows that $\frac{1}{6} < I(y) < 
\frac{1}{3}$. 
It should be noted that $s$ is entirely fixed by the parameters of the 
zero-temperature model, determined at low energy. 
Its value is then $50 \leq s \leq 120\mbox{ MeV/fm}^2$, higher than the 
estimates given by lattice computations without dynamical quarks 
\cite{iwasaki94}, but in agreement with the surface tension obtained by 
Burakovsky \cite{burakovsky96} within an effective model of QCD different from 
ours.

$F$ is stationary at a radius $r_0 = \frac{2s}{\Delta P}$, and takes the value
\begin{equation}
S_3 = \frac{16\pi}{3}\frac{s^3}{\Delta P^2} = \frac{16\pi}{3} \frac{\left[
\lambda f_\pi^3 I(y) \right]^3}{\Delta P^2}
\label{action1}
\end{equation}

Let us call $t_c$ the time at which the temperature equals $T_c$. 
When $t \gtrsim t_c$, both phases with $\xi = 0$ and $\xi = f_\pi$ coexist. 
If $x(t)$ is the fraction of the Universe occupied by the physical vacuum, and
therefore $1 - x(t)$ the fraction taken up by the quark plasma, the energy
density of the matter which fills the Universe is
\begin{eqnarray}
\varepsilon(t) & = & [1-x(t)]\varepsilon_{\xi=0}(t) + x(t)\varepsilon_{\xi =
f_\pi}(t) \nonumber\\ 
 & = & [1-x(t)]\left[\varepsilon_q(t) + \varepsilon_v(t)\right] +
 x(t)\varepsilon_v(t) 
\label{en1}
\end{eqnarray}
$\varepsilon_q$, $\varepsilon_v$, $\varepsilon_{\xi = 0}$ and 
$\varepsilon_{\xi = f_\pi}$ stand respectively for the energy densities of the 
quark-gluon plasma, the thermal bath of photons and leptons, the metastable 
vacuum and the physical vacuum. 
They are given by
\begin{eqnarray}
 & \varepsilon_q  =  B - \gamma_T T^2 + 3\alpha_T T^4 & 
\label{en1b1}\\ 
 & \varepsilon_v = 3\alpha_v T^4  - \gamma_v T^2 & \label{en1b2}\\
 & \varepsilon_{\xi=0} = \varepsilon_q + \varepsilon_v  & \nonumber\\
 & \varepsilon_{\xi=f_\pi} = \varepsilon_v  & \nonumber
\end{eqnarray}
Expression (\ref{en1}) can be rewritten
\begin{equation}
\varepsilon(t) = [1-x(t)]\varepsilon_q(t) +  \varepsilon_v(t)
\label{en1t}
\end{equation}
It appears as the sum of two contributions corresponding to the quark plasma, 
which is going to disappear during the phase transition, and to the background 
of non-interacting particles. 
If we replace the energy densities $\varepsilon_q(t)$ and $\varepsilon_v(t)$ 
with their temperature-dependent expressions (\ref{en1b1}, \ref{en1b2}), we 
obtain
\begin{equation}
\varepsilon(t) = [1-x(t)]\left[B - \gamma_T T(t)^2 + 3\alpha_T T(t)^4\right]
+ 3\alpha_v T(t)^4 - \gamma_v T(t)^2
\label{en1q}
\end{equation}

If we now suppose that the nucleation of the bubbles of phase with $\xi = 
f_\pi$ is isotropic enough, the Universe is still described by the
Robertson-Walker metric. 
We can then deduce from Friedmann's equation and the previous expression the 
equation which governs the expansion :
\numparts
\begin{eqnarray}
\lefteqn{\frac{1}{R(t)}\frac{dR}{dt}(t) = \sqrt{\frac{8\pi}{3
m_P^2}\varepsilon(t)}} \label{Einstein11} \\ 
\frac{dR}{Rdt}& = & \sqrt{\frac{8\pi}{3 m_P^2}\left[(1-x) \left(B - \gamma_T
T^2 + 3\alpha_T T^4 \right) + 3\alpha_v T^4 - \gamma_v T^2\right]}
\label{Einstein12} 
\end{eqnarray}
\endnumparts
where $m_P$ is the Planck mass while $x$ and $T$ are time-dependent.
 
In the beginning, few bubbles appear, and the temperature still decreases.
As time goes on, the size of bubbles increases, because of both the
propagation of the bubble walls and the expansion of the Universe. 
For example, if we consider a bubble of radius $r(t)$ which appeared at time
$t_i$, its growth is given by $\frac{dr}{dt} = v + r\frac{1}{R}\frac{dR}{dt}$, 
hence 
\begin{equation}
r(t) = v R(t) \int_{t_i}^t\frac{dt^\prime}{R(t^\prime)}
\label{radius}
\end{equation}
$v$ is the propagation speed of the walls, which we shall take equal to the
velocity of light in our calculations.

While the size of existing bubbles increases, new bubbles form. 
Thus the number density of nucleation sites at time $t$ is
\begin{equation}
N(t) = \int_{t_c}^t \left[1-x(t^\prime)\right] \Gamma(t^\prime)
\left[\frac{R(t^\prime)}{R(t)} \right]^3 dt^\prime 
\label{N_t}
\end{equation}
This expression shows that the number $N(t)[R(t)]^3$ of bubbles in a unit 
comoving volume which exist at $t$, i.e.\ those which appeared at any time 
$t^\prime < t$, is increasing. 
However, the number of bubbles per unit physical volume $N(t)$ may either 
increase, when the nucleation rate $\Gamma(t^\prime)$ is large, or decrease, 
when the expansion factor $\left[\frac{R(t)}{R(t^\prime)} \right]^3$ becomes 
large. 
In other words, the mean physical distance between nucleation sites may either 
decrease, in the first case, or increase as the Universe expands. 
Nonetheless, the fraction $x(t)$ of the Universe filled up with physical 
vacuum is growing with the number of bubbles. 
As we will see later, it is necessary to take into account the expansion of 
the Universe in the evolution of this fraction, so that 
\begin{equation}
x(t) = 1 - \exp\left(-\int_{t_c}^t \left[1-x(t^\prime)\right]
\Gamma(t^\prime) \frac{4\pi}{3} \left[v R(t)
\int_{t^\prime}^t\frac{dt^{\prime\prime}}{R(t^{\prime\prime})} \right]^3
dt^\prime \right)  
\label{x_t}
\end{equation}
This expression is analogous to the formula derived by Guth and Weinberg 
\cite{guth81b} in the context of GUT phase transitions, with the exception of 
the extra factor $1-x(t^\prime)$, which takes into account the fact that  
hadronic bubbles can only appear within the quark plasma. 

The evolution of $T$ during the phase transition is easily determined by
using the energy conservation, which reads
\begin{equation}
[\varepsilon(t) + P(t)] \frac{dR^3}{dt} + R(t)^3 \frac{d\varepsilon}{dt} = 0
\label{en_cons}
\end{equation}
$\varepsilon(t)$ is the energy density, given by expression (\ref{en1q}),
whereas the pressure is
\begin{equation}
P = (1-x)(\alpha_T T^4 - \gamma_T T^2 - B) + \alpha_v T^4 - \gamma_vT^2
\label{press}
\end{equation}
The time-dependences of $\varepsilon$ or $P$ are connected with those of $T$
and $x$, so we get
\begin{eqnarray}
\frac{dT}{dt} & = & -\frac{(1-x)\left(6 \alpha_T T^3 - 3 \gamma_T T \right)
+ 6\alpha_v T^3 - 3\gamma_v T}{(1-x) \left(6 \alpha_T T^2 - \gamma_T
\right) + 6 \alpha_v T^2 - \gamma_v} \frac{dR}{Rdt} \nonumber \\
 & + & \frac{3 \alpha_T T^4 - \gamma_T T^2 + B}{(1-x) \left(12 \alpha_T T^3
 - 2 \gamma_T T \right) + 12 \alpha_v T^3 - 2\gamma_v T} \frac{dx}{dt}
\label{dT_dt1}
\end{eqnarray}

$\frac{dT}{dt}$ is the sum of two terms : the first one is the usual 
contribution of the expansion of the Universe, which decreases the 
temperature. 
The second contribution comes from the replacement of the false vacuum $\xi = 
0$ with the physical vacuum $\xi = f_\pi$, which has a much lower specific 
heat, and corresponds to the release of latent heat, which tends to  increase 
$T$. 
In earlier studies \cite{witten84,fuller,iso86,ng91}, it is assumed that the latent 
heat released by the first bubbles of hadronic phase formed when the temperature 
drops under $T_c$ quickly reheats the Universe up to the critical temperature, 
and further bubble nucleation is suppressed. 
For the remainder of the transition, i.e. some microseconds, the temperature 
is kept constant by the latent heat due to the growth of the hadronic bubbles, 
which balances the expansion of the Universe. 
This latter remains negligible: equation (\ref{radius}) reduces to $r(t) \approx 
v(t-t_c)$, while the expansion factor in expression (\ref{N_t}) is almost equal to 1. 
In this work, we investigate this point more carefully, following the detailed 
evolution of the temperature. 
In particular, we will allow out-of-equilibrium coexistence at a temperature 
different from $T_c$ between the quark plasma and the hadronic phase. 
For example we can expect from equation (\ref{dT_dt1}) that the product
$R(t)T(t)$ will not remain constant during the transition, and the total
entropy of the Universe, which is proportional to $(RT)^3$, will not be
conserved.   

In many calculations concerning first order cosmological phase transitions
\cite{guth81b,guth81a}, it is assumed that $R(t)T(t) = \mbox{constant}$ 
during the transition, i.e.\ that entropy is conserved. 
In that case, the second term of equation (\ref{dT_dt1}) is neglected, and
following the work of Coleman \cite{coleman77} it is thought that all the
latent heat is taken by the bubble walls. 
At the completion of the transition the Universe would then reheat as the 
energy in the walls is dissipated by their collisions \cite{hawking82}, and $T$ 
would quickly increase up to a temperature close to its initial value $T_c$. 
In this work we will see that there is no such steep drop of temperature, but 
that the latent heat is released gradually, and holds $T$ rather high, though 
not necessarily at $T_c$, during the entire transition. 
When this latter is completed, the temperature may then be significantly 
smaller than the critical temperature $T_c$.

After a while, bubbles begin to coalesce and to enclose lumps of the phase
with $\xi = 0$. The quark number enclosed in such a chunk is conserved,
whereas the size of the lump decreases as the bubble walls propagate, and
therefore its density increases, until the pressure of the quark plasma
trapped inside balances the pressure of the physical vacuum $\xi = f_\pi$.
At that time, we can consider that $x \simeq 1$, because the regions filled
with quark plasma are but a minute fraction of the Universe, and the
transition has come to an end.
Let us call $t_f$ the time at which $x(t_f) = 1$, and $N(t_f)$ the 
corresponding number density of nucleation sites. 
To simplify the calculation we have not considered the dispersion in the 
number of trapped quarks in the chunks as given by the statistical 
distribution of nucleation sites. 
To estimate the mean number of quarks per body, we shall assume that these 
sites are on the vertices of a cubic lattice. 
At the centre of each cube is a quark clump. 
If no phase transition had occurred, the quarks enclosed would take up a 
volume $V_f = \frac{1}{N(t_f)}$, which corresponds to a physical volume $V_c = 
V_f\left[\frac{R(t_c)}{R(t_f)}\right]^3$ at the beginning of the transition. 
Therefore, the body contains ${\cal N}_q = n_q(t_c)V_c$ quarks, where $n_q(t)$ 
is the quark number density at time $t$, which can be easily related to the 
photon number density $n_\gamma = \frac{2}{\pi^2} \zeta(3) T^3$ and the quark 
to photon ratio $\frac{n_q}{n_\gamma}$. 
The conservation of the quark number when there is no baryon number violating 
interaction reads $n_q(t_c)[R(t_c)]^3 = n_q(t_f)[R(t_f)]^3$, so it is possible to 
rewrite ${\cal N}_q = n_q(t_f)V_f = \frac{n_q(t_f)}{N(t_f)}$, hence
\begin{equation}
{\cal N}_q = \frac{2}{\pi^2} \zeta(3) \frac{n_q}{n_\gamma}(t_f)
\frac{T_f^3}{N(t_f)}
\label{N_q}
\end{equation}
$\frac{n_q}{n_\gamma}(t_f)$ is the quark to photon ratio at the end of the
transition. We suppose that this value remained unchanged till the beginning
of the Big Bang Nucleosynthesis at $T \simeq 1\MeV$, and therefore take in
our calculations $\frac{n_q}{n_\gamma}  = \frac{3n_B}{n_\gamma}  \simeq
10^{-9}$, with $n_B$ the baryon number density \cite{walker91}. 
As a matter of fact, this value of $\frac{n_q}{n_\gamma}$ is valid in the 
hadronic phase, not in the quark plasma. 
However, estimates \cite{witten84,iso86} show that the baryon number density 
in the quark phase should be larger than in the hadron phase, and thus our 
assumption tends to underestimate ${\cal N}_q$.

\section{Results}

We have tested our model with different values of the parameters $B$, $m_s$
and $\lambda$. With a high value of $B$, such as $B^{1/4} = 200\MeV$, and
$m_s = 100\MeV$, the critical temperature is $T_c \simeq 148\MeV$. 
If we take $\lambda = 18$, we find the usual results \cite{witten84,ng91}, 
although our model is different, and the number of quarks trapped in a nugget 
at the end of the transition, which is completed in some microseconds, is 
${\cal N}_q \simeq 10^{40}$. 
In that case, the standard model of cosmology shows that the horizon at $t_f$ is 
too small to allow any object with a mass exceeding $10^{-8}\Msun$ to be 
formed \cite{wnc91,alam99}, which is in agreement with what we find. 
However, it may be worth noting that the upper limit given by this simple but 
powerful argument will no longer be valid if the Hubble volume is larger than 
what is given by the strictest standard model of cosmology (i.e. $R(t) \propto 
t^{1/2}$). 

Another limiting situation appears when $B$ takes a very low value, as for
example $B^{1/4} = 50\MeV$. 
In this case, we find that $x(t)$ cannot reach $1$, and therefore the 
transition never ends. 
The energy density which confines quarks within solitons is so weak that few
bubbles of the physical vacuum $\xi = f_\pi$ are nucleated, and they are so far
away from each other that their walls cannot meet while the Universe expands. 
An analogous phenomenon occurs when the surface tension $s$ is too high. 
This happen also if the mean distance between bubbles $[N(t_f)]^{1/3}$ is smaller 
i.e.\ if the value of $B$ is larger, but if simultaneously the speed of the bubble walls 
$v$ is decreased. 
Once again, the expansion of the bubbles cannot catch up with the expansion of 
the Universe.

Now if $B$ takes intermediate values, like $B^{1/4} = 120\MeV$ (i.e., in other 
units, $B = 0.61\mbox{ fm}^{-4}$, which is obtained from fits to hadronic properties 
in \cite{dethier86}) with $m_s = 100\MeV$, which gives a critical temperature 
$T_c = 90.1\MeV$, and $\lambda \simeq 17.77$, corresponding to a glueball 
mass of $1.6\GeV$, the number of quarks enclosed is ${\cal N}_q \simeq 
10^{57}$, i.e.\ a quark content of approximately $0.3\Msun$. 
We have plotted the evolution during the transition of several quantities in 
such a case: the fraction $x(t)$ of Universe filled with physical vacuum, the 
temperature $T(t)$ and the scale factor $R(t)$.

\begin{figure}
\begin{center}
\epsfbox{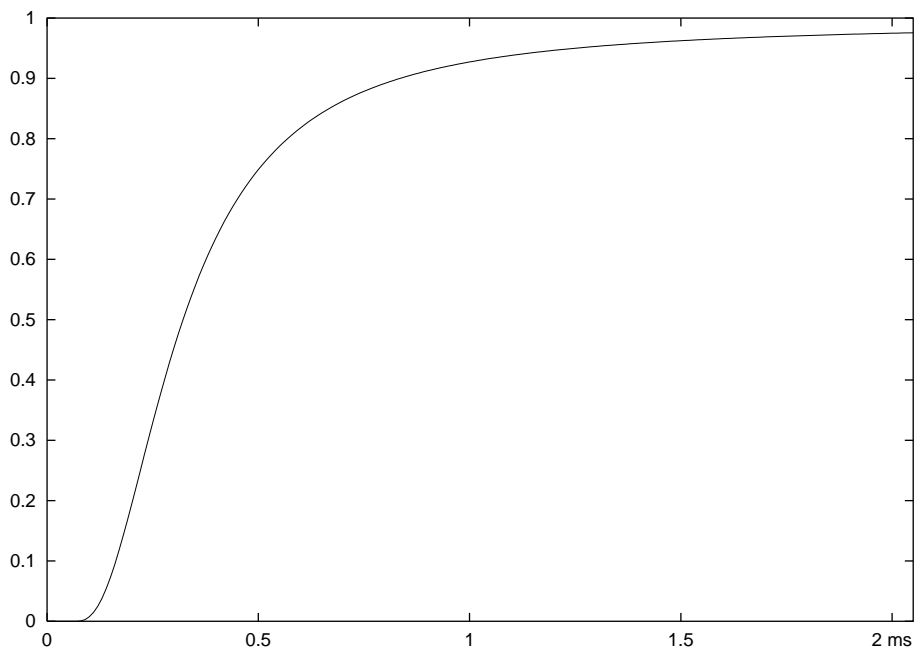}
\end{center}
\caption{Evolution of the fraction of the Universe filled up
with the  phase $\xi = f_\pi$ during the transition in the case 
$B = (120\MeV)^4$, $m_s = 100\MeV$, $\lambda \simeq 17.77$.\label{plot-x}} 
\end{figure}

We can see in \fref{plot-x} that $x(t)$ remains close to $0$ for some time, 
when few bubbles have yet appeared. 
Then it rises quickly, though not instantaneously as in the ``fast'' scenario, and 
tends towards its asymptotical value $1$.  
The time scale of the transition is now of the order of the millisecond, rather 
than the microsecond.

\begin{figure}
\begin{center}
\epsfbox{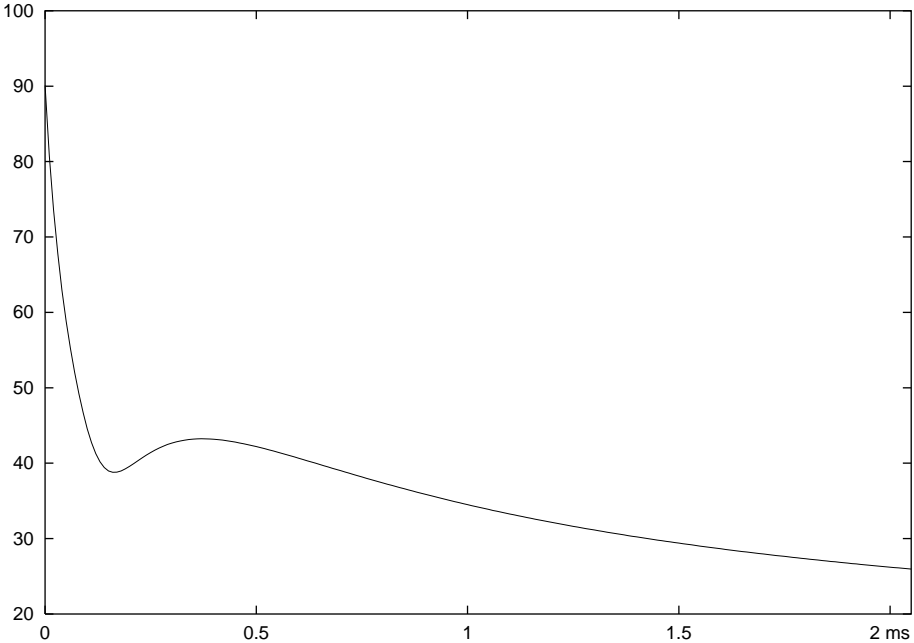}
\end{center}
\caption{Evolution of temperature during the transition.\label{plot-T}}
\end{figure}

The temperature (\fref{plot-T}) behaves as we had foretold. 
In the beginning, it decreases with a law close to the $t^{-1/2}$ law of the 
standard model: the first term of the r.h.s.\ of equation (\ref{dT_dt1}) is 
dominant. 
Then it increases, though not much, when one phase is replaced with another: 
the influence of the term in $\frac{dx}{dt}$, which represents the release of 
latent heat, is the more important;  we can check that the fast growth of $x$ 
and the increase in $T$ are simultaneous. 
Finally, $T$ decreases like $t^{-1/2}$ again, when the Universe is mostly 
filled up with physical vacuum, down to $T_f = 17.7\MeV$.

\begin{figure}
\begin{center}
\epsfbox{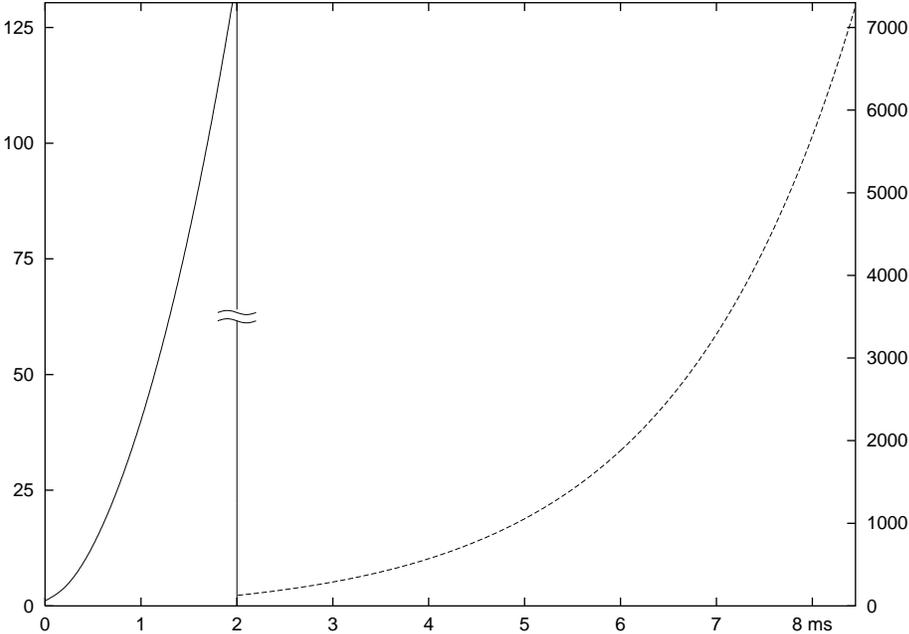}
\end{center}
\caption{Evolution of the scale factor during the transition.\label{plot-R}}
\end{figure}

The scale factor $R$ is plotted in \fref{plot-R}. 
Obviously, it does not follow the $t^{1/2}$ law of the standard model, but 
rather seems to grow exponentially, to end up with a value $R_f \simeq 7 
\times 10^3 R_c$. 
It should be noted that in this case, the dilution factor in equation (\ref{N_t}) is 
no longer negligible, since $\left[\frac{R(t^\prime)}{R(t)} \right]^3$ can be as 
small as $10^{-11}$! 
During the transition, the scale factor has been multiplied by $\simeq 10^4$, 
and therefore the Hubble radius is $10^4$ times the radius predicted in the 
strict standard model: the Hubble volume is $\simeq 10^{11}$ times larger than 
what is expected, and there is no contradiction between the number ${\cal N}_q$  
of quarks enclosed in a nugget and the total number of quarks within the Hubble 
volume.

If we compare our model with Guth's inflationary model \cite{guth81a}, we can 
trace whence such a growth comes. 
As in his case, when the temperature has dropped sufficiently, its contribution in 
our equations (eq. \ref{en1q}) can be neglected with regard to the metastable 
vacuum energy density $B$. 
This latter then drives the expansion \cite{patzelt89}. 
However, in our model the exponential growth comes to a natural end according 
to equation (\ref{Einstein12}), since the factor $1-x(t)$ vanishes when the 
transition is completed. 
A second difference between our model and old inflation \textit{\`a la} Guth 
regards entropy. 
In his model entropy is conserved during the exponential expansion and 
increases, due to reheating when bubbles collide, only at the completion of the 
transition. 
On the contrary, here entropy is constantly increasing during the quark-hadron 
phase transition, as the product $R(t)T(t)$ does not remain constant. 
To be more precise, the scale factor $R$ is multiplied by $7\times10^3$, whereas 
$T$ only decreases from $90$ to $18\MeV$: the product $RT$ is multiplied by 
more than $1.4 \times 10^3$, and hence the entropy increases by a factor $3 
\times 10^9$.

This increase in entropy has radical consequences on the evolution of the
ratio $\frac{n_q}{n_\gamma}$, which is a measure of the baryon asymmetry of
the Universe, during the transition. 
We have already seen that if there is no baryon number violating interaction, 
the quark number density $n_q$ decreases as $R^{-3}$, and that $n_\gamma 
\propto T^3$; hence, the ratio $\frac{n_q}{n_\gamma} \propto (RT)^{-3}$ is 
proportional to the inverse of the entropy. 
In particular, if, as we have found before, the entropy is multiplied by 
$10^9$ during the quark-hadron transition, the baryon asymmetry is divided by 
the same factor. 
Therefore, if at the end of the transition $\frac{n_q}{n_\gamma} \simeq 
10^{-9}$ as required by primordial nucleosynthesis, it means that at $T \gtrsim 
T_c$ the ratio was of the order $1$. 
The results found here contrast sharply with normal adiabatic expansion in 
which this baryon asymmetry would not change. 
However, when $\frac{n_q}{n_\gamma} \simeq 1$, the approximation made so far 
of neglecting the quark density or the chemical potential of the quarks with regard to 
$T$ is no longer valid.

\section{Finite chemical potential}

To handle the possibility of large quark densities, we must now incorporate
chemical potential into our model. 
Within the one loop approximation, the contributions of fermionic degrees of 
freedom which should be added to the effective potential $V(\xi)$ at zero 
temperature and zero chemical potential are no longer given by expression 
(\ref{free_en_ferm1}), but by
\begin{equation}
\fl \omega_f(T, \mu) = -T\left[\int\!\frac{d^3k}{(2\pi)^3} \ln\left(1 +
e^{-(E(k)-\mu)/T}\right) + \int\!\frac{d^3k}{(2\pi)^3} \ln\left(1 +
e^{-(E(k)+\mu)/T}\right)\right] \\
\label{free_en_ferm2}
\end{equation}
The first term is the fermion contribution, the second is from the 
antifermions. 
We have changed the notation from $f_f$ to $\omega_f$ because at finite 
$\mu$, the free energy and the thermodynamic potential do not coincide any 
more. 
For the same reason as before, the excitations of the scalar field vanish at 
both minima of $V(\xi)$, whereas those of quark-antiquark pairs are important 
only for $\xi$ close to $0$, where it becomes possible to expand the 
right-hand side of equation (\ref{free_en_ferm2}) as
\begin{equation}
\omega_f(T, \mu) = -\frac{7\pi^2}{360} T^4 -\frac{1}{12}\mu^2 T^2 -
\frac{1}{24\pi^2}\mu^4 + \frac{1}{24}g^2 \xi^2 T^2 + \frac{1}{8\pi^2}g^2
\xi^2 \mu^2 
\label{free_en_ferm2b}
\end{equation}
As previously, we keep the mass terms only for the strange quark. 
The self-interaction potential at $\xi = 0$  reads 
\begin{eqnarray}
 & V_{T, \mu}(\xi) = B - \alpha_T T^4 -\gamma_{\mu T}\mu^2 T^2 - \alpha_\mu
\mu^4 + \gamma_T T^2 + \gamma_\mu \mu^2  &  
\label{eff_pot_Tmu} \\
 & \alpha_T = \frac{7\pi^2}{20}, \quad 
\gamma_{\mu T} = \frac{3}{2}, \quad 
\alpha_\mu = \frac{3}{4\pi^2}, \quad 
\gamma_T = \frac{1}{4} m_s^2, \quad 
\gamma_\mu = \frac{3}{4\pi^2} m_s^2 & \nonumber
\end{eqnarray}
As before, we consider $B$ and $m_s$ as parameters, while $\mu$ is related
to the quark number density by $n_q = 4 \alpha_\mu \mu^3 + 2 
\gamma_{\mu T} \mu T^2 - 2 \gamma_\mu \mu$.

Therefore, the difference in pressure between the two minima of 
$V_{T, \mu}(\xi)$ is
\begin{equation}
\Delta P = B + \gamma_T T^2 + \gamma_\mu \mu^2 - \alpha_T T^4 - 
\gamma_{\mu T}\mu^2 T^2 - \alpha_\mu\mu^4
\label{diff_P2}
\end{equation}
This in turn changes the expression of the critical temperature, given by the 
condition $\Delta P = 0$, and of the bubble nucleation rate (\ref{rate}). 

The energy density which drives the expansion of the Universe through 
Friedmann's equation (\ref{Einstein11}) is now
\begin{equation}
\fl \varepsilon = (1-x)\left(B  -\gamma_T T^2 - \gamma_\mu \mu^2 + 3\alpha_T  T^4 +
3\gamma_{\mu T}\mu^2 T^2 + 3\alpha_\mu \mu^4 \right) + 3\alpha_v T^4 -
\gamma_v T^2
\label{epsilon2}
\end{equation}
whereas expression (\ref{dT_dt1}), which gives the evolution of temperature
during the transition, becomes 
\begin{equation}
\fl \eqalign{\frac{dT}{dt} & = -\frac{3T\left[(1-x)\left(2\alpha_T T^2 + 
\gamma_{\mu T}\mu^2 -\gamma_T -\frac{\gamma_{\mu T}\mu n_q}{6\alpha_\mu \mu^2 +
\gamma_{\mu T} T^2 - \gamma_\mu}\right) + 2\alpha_v T^2 - \gamma_v
\right]}{(1-x)\left(6\alpha_T T^2 + \gamma_{\mu T}\mu^2 -\gamma_T
-\frac{4\gamma_{\mu T}^2 \mu^2 T^2}{6\alpha_\mu \mu^2 + \gamma_{\mu T} T^2 -
\gamma_\mu}\right) + 6\alpha_v T^2 - \gamma_v}\frac{dR}{Rdt} \\ 
 & + \frac{B + 3\alpha_T T^4 + 3\alpha_\mu \mu^4 + 3\gamma_{\mu T}\mu^2
 T^2  - \gamma_T T^2 -\gamma_\mu \mu^2}{2T\left[(1-x)\left(6\alpha_T T^2 +
 \gamma_{\mu T}\mu^2 -\gamma_T -\frac{4\gamma_{\mu T}^2 \mu^2
 T^2}{6\alpha_\mu \mu^2 + \gamma_{\mu T} T^2 - \gamma_\mu}\right) + 6\alpha_v
 T^2 - \gamma_v T^2 \right]}\frac{dx}{dt}} \label{dT_dt2}
\end{equation}
The remainder of the calculation of the number ${\cal N}_q$ of quarks trapped 
in a typical chunk is changed accordingly. 
However, whereas the number of the model parameters $B$, $m_s$ and $\lambda$ 
remains the same as in the vanishing chemical potential case, there is now an 
important extra requirement, namely that the quark to photon ratio after the 
transition should be equal to $10^{-9}$, in agreement with the value required 
for primordial nucleosynthesis. 
From now on, we choose to take $B^{1/4} = 120\MeV$ and $m_s = 100\MeV$, 
whereas $\lambda$ will be determined by the above condition. 
In other words, once 2 of our 3 parameters have been fixed, for example the energy 
density of the false vacuum and the mass of the strange quark, the value of the third 
parameter cannot be chosen at will, but is dictated by the amount of entropy which has 
to be released during the transition. 
Hopefully, as we will see, the ``good'' values of $\lambda$ come exactly in the range 
which gives the glueball mass.

If we take a chemical potential $\mu(t_c) = 31.44\MeV$, equivalent to
$\frac{n_q}{n_\gamma}(t_c) = 1$ at the critical temperature $T_c = 89.96\MeV$, 
and with $\lambda \simeq 17.77$, at the end of the phase transition, the 
number of quarks enclosed in a chunk is ${\cal N}_q \simeq 2 \times 10^{57}$, 
i.e.\ approximately $0.5\mbox{ M}_\odot$. 
The final temperature is $T_f = 17.91\MeV$ and the ratio of the final and initial 
scale factors $\frac{R_f}{R_c} \simeq 4 \times 10^3$, so that the quark to photon 
ratio after the transition has become $\frac{n_q}{n_\gamma} \simeq 2 \times 
10^{-9}$. 
However, we have to wonder how an important quark number density could have 
arisen in the early Universe, at temperatures above the QCD scale. 

Affleck and Dine \cite{affleck85} have proposed a mechanism for baryogenesis
which could give such a high initial value of $\frac{n_q}{n_\gamma}$. 
In their model, the decay when supersymmetry breaking effects become 
important of the large vacuum expectation value of a scalar fermion, acquired 
either by quantum fluctuations at the Planck epoch or after an episode of inflation 
\cite{connors93}, leads  at a temperature of the order $10\mbox{ TeV}$ to a baryon 
asymmetry which can be as large as  $\frac{n_q}{n_\gamma}=10^3$. 
The actual value of the asymmetry depends on several parameters of the model, 
such as the expectation value of the squark field which decays or the
CP-violating phase required to produce more quarks than antiquarks.
After its production, this huge baryon asymmetry could still have been wiped out, 
long after the SUSY breaking epoch, by the so-called sphaleron transitions  
\cite{klinkhamer84} arising, for example, from the electroweak anomaly.
However, it was shown in reference \cite{morgan91} that a baryon-lepton number 
($B-L$)\footnote[1]{Throughout this paragraph, $B$ denotes the baryon number, 
not the energy density of the chirally symmetric vacuum $\xi = 0$} excess can be 
produced by the Affleck-Dine mechanism along with the usual $B$ excess. 
This is sufficient to prevent any electroweak sphaleron induced $B$ violation 
which would erase the above mentioned baryon asymmetry.
Furthermore, if we leave aside the electroweak symmetry breaking transition,
which, within the standard model, does not seem to affect much the baryon
asymmetry \cite{dine92}, and assuming normal adiabatic, radiation dominated
evolution, the quark to photon ratio left by the Affleck-Dine baryogenesis
should not be significantly changed down to the QCD transition considered 
here. 

Therefore, we have taken several different ratios $\frac{n_q}{n_\gamma}(t_c)$, 
keeping in mind the necessary requirement $\frac{n_q}{n_\gamma}(t_f) \simeq 
10^{-9}$ at the end of the transition.
This requirement can be met, for moderate values of the baryonasymmetry, by 
changing slightly the value of $\lambda$. 
However, for larger values of $\frac{n_q}{n_\gamma}(t_c)$ the change in 
$\lambda$ is more pronounced, in order to get enough expansion to dilute the 
initial baryon asymmetry to the desired value.
The results with $B^{1/4} = 120\MeV$, $m_s = 100\MeV$ and different values
of $\lambda$ are shown in \tref{results}. 

\begin{table}
\caption{Values of ${\cal N}_q$ with $B^{1/4} = 120\MeV$, $m_s =
100\MeV$, $\lambda$ variable.\label{results}} 

\begin{indented}
\lineup
\item[]\begin{tabular}{@{}lllllllll} 
\br
$\frac{n_q}{n_\gamma}(t_c)$&$\lambda$&$T_c$&$\mu(t_c)$&
$T_f$&$t_f$& $\frac{R_f}{R_c}$&$\frac{n_q}{n_\gamma}(t_f)$
& ${\cal N}_q$ \\
 & &(MeV)&(MeV)&(MeV)&(ms)& & & \\
\mr
\0$0.01$&$17.77$&$90.10$&\0$8.98$&$19.81$&$4.612$&\0$727$&$2\times10^{-9}$&
$1.8\times10^{55}$ \\ 
\0$0.1$&$17.77$&$90.10$&$15.48$&$18.56$&$5.990$&$1739$&$2\times10^{-9}$&
$1.8\times10^{56}$ \\ 
\0$1$&$17.77$&$89.96$&$31.44$&$17.91$&$7.391$&$3974$&$2\times10^{-9}$&
$1.8\times10^{57}$ \\ 
$10$&$17.56$&$87.28$&$64.94$&$17.60$&$8.706$&$8405$&$2\times10^{-9}$&
$1.5\times10^{58}$ \\ 
$20$&$17.27$&$83.70$&$78.41$&$17.64$&$8.636$&$10^4$&$2\times10^{-9}$&
$5.2\times10^{57}$ \\ 
$30$&$16.99$&$80.25$&$86.05$&$17.83$&$7.970$&$10^4$&$2\times10^{-9}$&
$1.6\times10^{57}$ \\ 
$40$&$16.73$&$77.09$&$90.98$&$18.78$&$6.190$&$5523$&$2\times10^{-8}$&
$3.1\times10^{56}$ \\
\br
\end{tabular}
\end{indented}
\end{table}

The first result which is striking in this table is the apparent proportionality between 
the initial quark number asymmetry $\frac{n_q}{n_\gamma}(t_c)$ and the number of 
quarks enclosed in a typical quark star. 
When $\frac{n_q}{n_\gamma}(t_c)$ grows from $0.01$ to $10$, ${\cal N}_q$ takes 
values corresponding to a quark content of $0.005$ to $5\Msun$. 
This seems natural: let us assume that the parameters $B$ and $m_s$ (remember that 
$\lambda$ is no longer a free parameter) determine the geometrical repartition of bubbles, 
i.e.\ the mean distance between nucleation sites. 
Then, the more quarks you put initially in the volume which corresponds to a future 
quark object, the more quarks you will have in that object in the end. 
However, when $\frac{n_q}{n_\gamma}(t_c)$ is larger than $20$, the number of quarks 
${\cal N}_q$ does not increase anymore, but on the contrary it decreases. 
This means that it is impossible to form quark stars with gigantic masses, even with fine 
tuning, whereas in previous studies such fermion soliton stars could have a mass as high 
as $10^{12}\Msun$ \cite{lee92,lee87}.

So far, we have assumed that the quark-hadron phase transition takes place
between the quark phase and the QCD vacuum. 
In a subsequent study, we have considered the more realistic case with the
transition between the quark  plasma and a hadron phase containing the 
lightest hadrons, the pions. 
These contribute only to the phase $\xi = f_\pi$, by an additional term
$-\frac{\pi^2}{30}T^4 + \frac{m_\pi^2}{8}T^2$ to the self-interaction 
potential $V_{T,\mu}(\xi=f_\pi)$, which changes the expression of the
overpressure (eq.\ \ref{diff_P2}) and the value of the critical temperature.
Similarly, a term $x(\frac{\pi^2}{10}T^4 - \frac{m^2_\pi}{8}T^2)$ must be
added to the energy density (eq.\ \ref{epsilon2}), thereby modifying 
Friedmann's equation and the energy conservation (eq.\ \ref{dT_dt2}).
The resulting calculations show that the qualitative features presented above 
remain unchanged, provided we slightly tune the value of $\lambda$.  
For example, with $B^{1/4} = 120\MeV$, $m_s = 100\MeV$ and an initial 
baryon asymmetry $\frac{n_q}{n_\gamma}(t_c)=10$, the baryon asymmetry 
at the end of the transition equals $10^{-9}$ if we take $\lambda = 17.76$, and 
the number of quarks enclosed in a typical quark star is ${\cal N}_q = 1.7 \times 
10^{58}$, close to the value obtained when pions are neglected.

\section{Conclusion}

We have studied the quark-hadron phase transition within an effective model
of QCD, the chiral quark model,  including finite temperature and chemical 
potential. 
We would like to stress that this type of effective models is widely used in 
various fields of physics like the low energy hadron physics or relativistic 
heavy ion collisions. 
It is also worth noting that the so-called inflaton or dilaton scalar field 
models often used in cosmology are special cases of this model. 
In our work, all the parameters of the model are fixed by fits to the static 
properties of hadrons, and we have seen that in a reasonable range of the key 
parameters $B$, $m_s$ and $\lambda$, quark plasma objects with very 
different masses can have been formed at the end of the transition, which are 
likely to survive until our epoch if their quark content is ${\cal N}_q \gtrsim 
10^{40}$ (see the discussion and the references in \cite{alam99}). 
The most interesting case regards the formation of a body with stellar mass, 
at a temperature much lower than the critical temperature of the transition. 
Such quark stars could be identified with some of the dark bodies detected by 
microlensing experiments in the halo of our galaxy \cite{MACHOs}.  
A new feature of our calculations, independent of the size of the formed quark 
bodies, is the presence of a period of exponential expansion of the Universe 
during the transition which ends in a natural way with the transition. 
This growth, unexpected at such a temperature, could account for the observed 
baryon asymmetry in the Universe, if a high value of the quark to photon ratio 
had been left before the transition, at $T \gtrsim 100\MeV$  by a baryogenesis 
mechanism \textit{\`a la} Affleck-Dine, i.e.\ if we drop the usual assumption 
that fermion densities in the (very) early Universe were negligible: the 
entropy production by the released latent heat could have diluted the 
asymmetry down to the value it must have at $T \lesssim 10\MeV$ to agree with 
the standard Big Bang nucleosynthesis. 
Actually, various models of entropy production have been designed for this purpose 
at very high energies by several authors \cite{ellis87}, but our dilution mechanism 
is different from those earlier proposals in that it involves only the physics at 
the QCD scale.


\Figures

\end{document}